\documentclass[12pt,preprint]{aastex}

\usepackage{longtable}
\usepackage{amsmath}
\renewcommand{\thefootnote}{\fnsymbol{footnote}}

\begin{document}

\title{Chemical Analysis of a Diffuse Cloud along a Line of Sight Toward W51: Molecular Fraction and Cosmic-Ray Ionization Rate \footnote{{\it Herschel} is an ESA space observatory with science instruments provided by European-led Principal Investigator consortia and with important participation from NASA}}

\author{Nick Indriolo\altaffilmark{1},
D.~A.~Neufeld\altaffilmark{1},
M.~Gerin\altaffilmark{2},
T.~R.~Geballe\altaffilmark{3},
J.~H.~Black\altaffilmark{4},
K.~M.~Menten\altaffilmark{5},
J.~R.~Goicoechea\altaffilmark{6}
}

\altaffiltext{1}{Department of Physics and Astronomy, Johns Hopkins University, Baltimore, MD 21218}
\altaffiltext{2}{LERMA, CNRS, Observatoire de Paris and ENS, France}
\altaffiltext{3}{Gemini Observatory, Hilo, HI}
\altaffiltext{4}{Department of Earth and Space Sciences, Chalmers University of Technology, Onsala Space Observatory, 43992 Onsala, Sweden}
\altaffiltext{5}{MPI f\"{u}r Radioastronomie, Bonn, Germany}
\altaffiltext{6}{Departamento de Astrof\'{i}sica, Centro de Astrobiolog\'{i}a (CSIC-INTA), 28850, Madrid, Spain}

\begin{abstract}

Absorption lines from the molecules OH$^+$, H$_2$O$^+$, and H$_3^+$ have been observed in a diffuse molecular cloud along a line of sight near W51 IRS2.  We present the first chemical analysis that combines the information provided by all three of these species.  Together, OH$^+$ and H$_2$O$^+$ are used to determine the molecular hydrogen fraction in the outskirts of the observed cloud, as well as the cosmic-ray ionization rate of atomic hydrogen.  H$_3^+$ is used to infer the cosmic-ray ionization rate of H$_2$ in the molecular interior of the cloud, which we find to be $\zeta_2=(4.8\pm3.4)\times10^{-16}$~s$^{-1}$.  Combining the results from all three species we find an efficiency factor---defined as the ratio of the formation rate of OH$^+$ to the cosmic-ray ionization rate of H---of $\epsilon=0.07\pm0.04$, much lower than predicted by chemical models.  This is an important step in the future use of OH$^+$ and H$_2$O$^+$ on their own as tracers of the cosmic-ray ionization rate.

\end{abstract}

\keywords{astrochemistry -- cosmic rays -- ISM: molecules}

\section{INTRODUCTION} \label{section_intro}

\setcounter{footnote}{6}
\renewcommand{\thefootnote}{\arabic{footnote}}

In the past decade, H$_3^+$ has widely become regarded as an excellent tracer of the cosmic-ray ionization rate in diffuse molecular clouds.  Surveys of H$_3^+$ in such clouds \citep{indriolo2007,indriolo2012} have enabled us to find variations in the ionization rate between sight lines, and to build up the distribution function of cosmic-ray ionization rates in the nearby interstellar medium (ISM).  However, observations of H$_3^+$ are currently limited to background sources with $L$-band magnitudes brighter than about $L=7.5$~mag.  At this cutoff, OB stars are only feasible as background sources to distances of a few kpc, meaning that H$_3^+$ observations are primarily limited to the local spiral arm \citep[observations toward the Galactic center, e.g., ][use dust-embedded objects]{goto2002,goto2008,goto2011,oka2005,geballe2010}.  An alternative method for inferring the ionization rate utilizes the chemistry associated with the formation and destruction of OH$^+$ and H$_2$O$^+$ \citep{gerin2010,neufeld2010}, thought to be dependent primarily on hydrogen abstraction reactions with H$_2$ and dissociative recombination with electrons.  The HIFI instrument \citep{degraauw2010} aboard {\em Herschel} \citep{pilbratt2010} has provided the first opportunity to observe both OH$^+$ and H$_2$O$^+$ with very high spectral resolution, thus allowing the use of these ions in constraining the cosmic-ray ionization rate.  Background sources bright enough for THz spectroscopy are widely distributed throughout the Galaxy, and targets from the PRISMAS\footnote{PRobing InterStellar Molecules with Absorption line Studies} key programme range in distance from about 1~kpc to 12~kpc.  However, the ionization rate inferred from the oxygen chemistry is dependent upon an efficiency factor, $\epsilon$, at which atomic hydrogen ionized by cosmic rays will eventually be converted into OH$^+$.  In order to determine $\epsilon$, we present observations of OH$^+$, H$_2$O$^+$, and H$_3^+$ in sight lines toward W51 and compare the ionization rates inferred separately from the hydrogen chemistry and oxygen chemistry.

\subsection{Hydrogen Chemistry}

The interstellar chemistry of H$_3^+$ is rather simple, and the reactions surrounding this molecule are given in the top portion of Table \ref{tbl_reactions}.  H$_3^+$ is formed in a two-step process, beginning with the ionization of H$_2$ by cosmic rays, and quickly followed by a reaction of H$_2^+$ with H$_2$.  Some H$_2^+$ is destroyed by dissociative recombination with electrons or by charge transfer to atomic hydrogen, but these reactions are generally slow compared to the $\mathrm{H_2^+ + H_2}$ process.\footnote{This is no longer true for the $\mathrm{H_2^+ + H}$ reaction at low molecular fraction.}  Cosmic-ray ionization is the rate-limiting step in this process as it is many orders of magnitude slower than proton transfer from H$_2^+$ to H$_2$, and can be taken as the formation rate of H$_3^+$.  The primary destruction mechanisms for H$_3^+$ are dependent on the environment under consideration.  In diffuse molecular clouds, H$_3^+$ is predominantly destroyed via dissociative recombination with electrons.  In dense clouds, however, where the electron fraction is much lower, H$_3^+$ is destroyed by proton transfer to neutrals such as CO and O. 
{


\subsection{Oxygen Chemistry}

Reactions involved in the chemistry surrounding OH$^+$ and H$_2$O$^+$ are presented in the bottom portion of Table \ref{tbl_reactions}.  The formation of OH$^+$ begins with the ionization of atomic hydrogen by cosmic rays.  This is followed by endothermic charge transfer to oxygen to form O$^+$---a process highly dependent upon the relative populations in the fine structure levels of atomic oxygen \citep{stancil1999}---and hydrogen abstraction from H$_2$ to form OH$^+$.  OH$^+$ is either destroyed by further hydrogen abstraction to form H$_2$O$^+$, or by dissociative recombination with electrons.  The same is true for H$_2$O$^+$, but H$_3$O$^+$ is only destroyed by dissociative recombination with electrons.  A steady-state analysis of these reactions is employed in Section \ref{section_analysis} in inferring the ionization rate of atomic hydrogen and molecular hydrogen fraction from OH$^+$ and H$_2$O$^+$ abundances.

An alternative means of forming OH$^+$ is the reaction of O with H$_3^+$.  This process requires a high molecular hydrogen fraction---such that H$_3^+$ is formed efficiently from cosmic-ray ionization of H$_2$---and a low electron fraction---such that H$_3^+$ is predominantly destroyed by proton transfer to O, forming OH$^+$.  As we will show in Section \ref{section_analysis} that the OH$^+$ and H$_2$O$^+$ probed by our observations reside in gas with a low molecular hydrogen fraction, we omit this pathway from our analysis.

\section{METHODS}

\subsection{Target Characteristics}

Observations in the near to mid infrared at UKIRT (United Kingdom Infrared Telescope) and Gemini South were made using the embedded cluster W51~IRS2 ($\alpha$=19$^h$23$^m$40.0$^s$, $\delta$=+14$^{\circ}$31\arcmin 06\arcsec; J2000.0) as a background source.  THz observations of W51 in the PRISMAS key program were pointed toward the 3.6~cm continuum peaks e$_1$ and e$_2$ \citep{mehringer1994} at $\alpha$=19$^h$23$^m$43.9$^s$, $\delta$=+14$^{\circ}$30\arcmin 30\farcs5 (J2000.0), meaning there is a 65\arcsec\ separation between the IR and THz pointings.

Previous studies of the W51 star forming complex and giant molecular cloud \citep[e.g.,][]{carpenter1998,okumura2001,bieging2010,kang2010} show emission in CO at velocities between about 50~km~s$^{-1}$ and 70~km~s$^{-1}$.  This gas is thought to be dense and associated with the W51 region.  However, \citet{carpenter1998} also report a cloud in the solar neighborhood at 7~km~s$^{-1}$, and some weaker features at 15--25~km~s$^{-1}$, which are thought to arise from a diffuse molecular cloud and a primarily atomic cloud, respectively\footnote{These clouds are not reported by \citet{okumura2001}, \citet{bieging2010}, and \citet{kang2010} because they are outside of the covered velocity ranges.} \citep{neufeld2002,sonnentrucker2010}.  All of these velocity components are also seen in H~\textsc{i} 21~cm observations \citep[][see G49.5--0.4e spectrum]{koo1997}.  The 25~km~s$^{-1}$ component is much stronger (with respect to other components) in H than CO, supporting the conjecture that this is a primarily atomic cloud.  In the present study we are concerned primarily with the conditions of diffuse molecular clouds, and so will focus mainly on the absorption at 5--7~km~s$^{-1}$.  This material shows a few closely spaced absorption components \citep{sonnentrucker2010,godard2010} and is estimated to be at a distance of about 500~pc using a simple Galactic rotation curve analysis \citep{gerin2011}, but may be as close as 100--200~pc based on maps of the nearby neutral ISM \citep{welsh2010}.  At 500~pc the on-sky separation of 65\arcsec\ between the IR and THz pointings corresponds to a physical separation of 0.16~pc, so all observations of the diffuse molecular cloud(s) of interest should probe roughly the same material.

\subsection{Observations}

Observations made at UKIRT utilized CGS4 \citep[Cooled Grating Spectrometer 4;][]{mountain1990} with its echelle grating, $\sim$0\farcs4 wide slit, long camera, and $3\times2$ pixel sampling mode to yield a resolving power of about 37,000 (resolution $\sim$8~km~s$^{-1}$), in combination with a circular variable filter (CVF) to select the correct order.  Targets were nodded along the slit in an ABBA pattern to facilitate the removal of sky background.  The $R(1,1)^u$ and $R(1,0)$ transitions of H$_3^+$ at 3.668083~$\mu$m and 3.668516~$\mu$m, respectively, were targeted toward W51 IRS2 and $\alpha$~Lyr (observed for the purpose of removing telluric lines from the science target spectrum) on 2001 May 26.  Total integration time on the science target was 33.6 min.  Observations targeting the $R(0)$ through $R(3)$ transitions of the $v=1$--$0$ band of $^{12}$CO near 4.64~$\mu$m toward W51 IRS2 and the telluric standard $\alpha$~Aql were made on 2001 May 28 with an integration time of 9.6 min on the science target.

At Gemini South the {\em Phoenix} spectrometer \citep{hinkle2003} was used with its echelle grating and 0\farcs17 slit to produce a resolving power of about 70,000 (resolution $\sim$5~km~s$^{-1}$).  The L2734 filter was employed to select the order containing the $R(1,1)^l$ transition of H$_3^+$ at 3.715479~$\mu$m.  Observations targeting this transition toward W51 IRS2 and the telluric standard $\alpha$~Aql were made on 2010 Jul 23 with a total integration time of 4 min on the science target.  



The HIFI spectrometer aboard  {\it Herschel} was used to observe the $N=1-0$, $J=2-1$, transition of OH$^+$ and the $N_{K_{a}K_{c}}=1_{11}-0_{00}$, $J=3/2-1/2$ transition of {\it ortho}-H$_2$O$^+$, for which the strongest hyperfine components are at 971.804 GHz \citep{muller2005} and 1115.204 GHz \citep{murtz1998} respectively.  The observations were carried out in dual beam switch (DBS) mode, with the reference beams located 3$^\prime$ on either side of the source.  We used multiple local oscillator (LO) frequencies, separated by a small offset, to confirm the assignment of any observed spectral feature to either the upper or lower sideband of the (double side band) HIFI receivers.  For the H$_2$O$^+$ line, the observations were performed on 
2010 Oct 29 with 3 separate LO settings in the lower sideband of mixer band 5a (Observation Identifications [ObsIDs] 1342207693, 1342207694, and 1342207695).  For the OH$^+$ line, observations were carried out with three separate LO settings in the lower sideband of mixer band 4a on 2010 Oct 28 (ObsIDs 1342207642, 1342207643, and 1342207644).  

\subsection{Data Reduction}

Starting from raw data frames, our reduction procedure for IR data utilized standarad IRAF\footnote{http://iraf.noao.edu/} routines commonly used in spectroscopic data reduction.  Upon extracting one-dimensional spectra, data were transfered to Igor Pro\footnote{http://www.wavemetrics.com/} where we have macros written to complete the reduction \citep{mccall2001}.  A full description of the data reduction procedure---applicable to both H$_3^+$ and CO data---is presented in \citet{indriolo2011}.

{\it Herschel} data were processed to Level 2 in HIPE\footnote{Herschel Interactive Processing Environment} using the standard HIFI pipeline, providing fully calibrated spectra with the intensities expressed as antenna temperature and the frequencies in the frame of the Local Standard of Rest (LSR).  For each of the target lines, the signals measured in the two orthogonal polarizations were in excellent agreement, as were spectra obtained at the various LO settings when assigned to the expected sideband.  We combined the data from the multiple LO settings, and from both polarizations, to obtain an average spectrum for each line. 

\section{RESULTS}

Reduced spectra are shown in Figure \ref{fig_spectra}.  All species show multiple absorption features resulting from multiple gas clouds along the line of sight.  H$_3^+$ and CO have components at about 5~km~s$^{-1}$, 50~km~s$^{-1}$, and 65~km~s$^{-1}$, and the
5~km~s$^{-1}$ component corresponds to the diffuse cloud of interest in our study.  OH$^+$ and H$_2$O$^+$ show absorption over a wider range of velocities, but the analysis of these spectra is complicated due to hyperfine splitting.  The green curves show the absorption due to only the strongest hyperfine component (found as discussed in Section \ref{subsection_extract}), revealing that OH$^+$ and H$_2$O$^+$ exhibit absorption at the same velocities as H$_3^+$ and CO.  Additionally, OH$^+$ shows absorption at 25~km~s$^{-1}$ where HF has been observed \citep{sonnentrucker2010}.  While our fit to the H$_2$O$^+$ spectrum does not require a component at 25~km~s$^{-1}$, \citet{wyrowski2010} reported H$_2$O$^+$ absoprtion at 22.5~km~s$^{-1}$ toward W51 at a position 60\arcsec\ away from the PRISMAS pointing (17\arcsec\ from the IR pointing).  Most likely this difference is caused by the the lower signal-to-noise ratio in the PRISMAS spectrum compared to the WISH (Water In Star-forming regions with {\it Herschel}) spectrum.  Note, however, that for OH$^+$ and H$_2$O$^+$ the many components used to fit absorption features do not necessarily correspond to physical clouds, but are simply used as a means to quantify the optical depth as a function of velocity.  A more complete description of the spectra in Figure \ref{fig_spectra} is given in the figure caption.

\section{ANALYSIS} \label{section_analysis}

\subsection{Extraction of Column Densities} \label{subsection_extract}

Absorption features due to H$_3^+$ were fit with Gaussian functions using the procedure described in \citet{indriolo2012} in order to determine equivalent widths, velocity FWHM, and interstellar gas velocities.  All three of the absorption features in the $R(1,1)^l$ spectrum were fit simultaneously, and the resulting FWHM and gas velocities were used to aid in the simultaneous fitting of the six absorption features (three $R(1,1)^u$ and three $R(1,0)$ lines) in the $R(1,1)^u$ and $R(1,0)$ spectrum.  Absorption line parameters and column densities determined from this analysis are reported in Table \ref{tbl_IRabsorption}}.  Note that the severe blending of the $R(1,1)^u$ line at $\sim50$~km~s$^{-1}$ with the $R(1,0)$ line at $\sim5$~km~s$^{-1}$ introduces additional uncertainty to the parameters extracted for these lines.

In the present study we are only concerned with the diffuse foreground cloud, and the feature of interest in the CO spectra is the relatively weak absorption near 5~km~s$^{-1}$ that is seen in the $R(0)$ and $R(1)$ lines.  In the bottom panel of Figure \ref{fig_spectra}---noticeable in the $R(0)$, $R(2)$, and $R(3)$ spectra---it is seen that the continuum level slopes downward from about $-10$~km~s$^{-1}$ to $35$~km~s$^{-1}$.  It is unclear whether this feature is due to a broad outflow component or some artifact of the instrument or data reduction, but it hinders the determination of the continuum level across the absorption feature of interest, and thus the determination of the equivalent width.  
To remove the contribution to absorption from the dense line-of-sight gas, remove the sloping continuum level, and extract the equivalent width from the feature of interest, we fit each spectrum with the sum of five Gaussian components.  Three narrow components between about 30~km~s$^{-1}$ and 80~km~s$^{-1}$ fit the dense cloud absorption; one broad component centered near 20~km~s$^{-1}$ with a FWHM of $\sim60$~km~s$^{-1}$ removes the sloping continuum level; and one narrow component fits the absorption from the diffuse cloud at 5~km~s$^{-1}$.  Extracted line parameters and column densities in the optically thin approximation for the diffuse cloud are given in Table \ref{tbl_IRabsorption}.  In the case of the $R(2)$ and $R(3)$ lines only the first four components are used in the fit, and upper limits on the equivalent width are determined from the standard deviation on the continuum level after dividing the spectrum by the fit.

The spectra of OH$^+$ and H$_2$O$^+$ were fit using a procedure similar to that described by \citet{neufeld2010}.  Multiple Gaussian components with adjustable line-center optical depths, velocity dispersions, and line centroids were convolved with the hyperfine structure for each transition to obtain an optimal fit to the observed spectra.  From this analysis, we find $dN/dv$ (column density per unit velocity interval) as a function of LSR velocity, such that integrating the function between any two velocities provides the column density in that interval.  Results from this procedure are presented in Table \ref{tbl_THzabsorption}.

Total column densities (i.e., the sum of column densities over all rotational levels) in the diffuse foreground cloud at $\sim5$~km~s$^{-1}$ for molecules studied in this paper, along with species reported in the literature that are necessary for our analysis, are given in Table \ref{tbl_allmol}.  

\subsection{Cloud Conditions from CO}

Observations of CO are useful in constraining physical conditions within a cloud.  The ratio of column densities in the $J=1$ and $J=0$ levels gives an excitation temperature of 4.3~K, much lower than the expected kinetic temperature of $\sim70$~K, but higher than the temperature of the cosmic microwave background radiation.  This excitation temperature suggests that both radiative and collisional excitation play a role in exciting the $J=1$ state.
The observed excitation temperature for $J=0$ and $J=1$ implies that the density of the collision partners H$_2$ and H must be significantly below the critical density for the $J=1\rightarrow0$ transition of CO.  An analysis of the level populations that includes radiative and collisional excitation along with optical depth effects \citep[based on that in][]{neufeld1995} suggests a density of $n_{\rm H}\approx100$~cm$^{-3}$ (where $n_{\rm H}\equiv n({\rm H})+2n({\rm H}_2)$), in good agreement with the range of values found by \citet{godard2010}.

The fraction of carbon in the form of CO is also important in understanding the chemical conditions in the ISM.  Table \ref{tbl_allmol} shows that the relative abundance of CO with respect to total hydrogen is $8.01\times10^{-7}$, while that of C$^+$ is $1.14\times10^{-4}$ \citep[][in press]{gerin2012}, implying that carbon is primarily in ionized form.  If electrons are predominantly the result of singly ionized carbon, then the fractional electron abundance, $x_e\equiv n_e/n_{\rm H}$, can be approximated by $x({\rm C}^+)$.  Because C$^+$ ions (and thus electrons) are 140 times more abundant than CO, and because the rate coefficient for proton transfer from H$_3^+$ to CO is about 100 times slower than that for dissociative recombination of H$_3^+$ with electrons, destruction of H$_3^+$ by CO is negligible.

\subsection{Ionization Rate from H$_3^+$}

The standard steady-state analysis for the formation and destruction of H$_3^+$ in diffuse clouds gives
\begin{equation}
\zeta_{2}n({\rm H}_2)=k({\rm H}_3^+|e^-)n({\rm H}_3^+)n_e,
\label{eq_H3+_steadystate}
\end{equation}
where $\zeta_2$ is the ionization rate of H$_2$, $n({\rm X})$ is the number density of species X, $n_e$ is the electron density, and $k({\rm X|Y})$ is the rate coefficient for the reaction between species X and Y.  Equation (\ref{eq_H3+_steadystate}) can be re-arranged and substitutions made such that
\begin{equation}
\zeta_2=k({\rm H}_3^+|e^-)x_{e}n_{\rm H}\frac{N({\rm H}_3^+)}{N({\rm H}_2)},
\label{eqzeta2}
\end{equation}
as described in \citet{indriolo2012}, where $N({\rm X})$ is the column density of species X.  Variables on the right-hand side of equation (\ref{eqzeta2}) are determined as follows.  The molecular hydrogen column density, $N({\rm H}_2)$, is estimated from $N({\rm CH})$ \citep[][see Table \ref{tbl_allmol} herein]{gerin2010ch} using the empirical relationship $N({\rm CH})/N({\rm H}_2)=3.5^{+2.1}_{-1.4}\times10^{-8}$ from \citet{sheffer2008}.  The H$_3^+$-electron recombination rate coefficient, $k({\rm H}_3^+|e^-)$, has been measured in multiple laboratory experiments \citep[e.g.,][]{mccall2004,kreckel2005,kreckel2010} with consistent results, and we adopt the analytical expression from \citet{mccall2004} shown in Table \ref{tbl_reactions}.  As mentioned above, a hydrogen density of $n_{\rm H}=100$~cm$^{-3}$ is adopted from the CO analysis, and the electron fraction is approximated by $x({\rm C}^+)=1.14\times10^{-4}$.  Note, however, that this may underestimate $x_e$ in regions where the cosmic-ray ionization rate is high enough that the abundance of H$^+$ rivals that of C$^+$.  An estimate of where this occurs is given by the model chemistry in \citet{hollenbach2012}. In their model with $n_{\rm H}=100$~cm$^{-3}$ and $\zeta_2=4.6\times10^{-16}$~s$^{-1}$ the resulting electron fraction is about $2x({\rm C}^+)$ in regions of low molecular fraction ($f_{\rm H_2}\leq0.2$), but comparable to $x({\rm C}^+)$ in regions of higher molecular fraction where H$_3^+$ is expected to primarily form.  As such, we omit this effect from our analysis of the H$_3^+$ chemistry.  Lastly, $N({\rm H}_3^+)$ is determined from the observations presented herein, and is given in Table \ref{tbl_allmol}.  Using all of these values and assuming $T=70$~K---an average spin temperature found from H$_2$ observations \citep{savage1977,rachford2002,rachford2009}---we find a cosmic-ray ionization rate of $\zeta_2=(4.8\pm3.4)\times10^{-16}$~s$^{-1}$ in the diffuse cloud component at 5~km~s$^{-1}$.  The uncertainty in this value is primarily due to the scatter in the CH/H$_2$ relation, and an assumed 50\% uncertainty in the density.

\subsection{Molecular Hydrogen Fraction from OH$^+$ and H$_2$O$^+$}

A steady state analysis of the H$_2$O$^+$ abundance gives the equation
\begin{equation}
n({\rm OH}^+)n({\rm H}_2)k({\rm OH}^+|{\rm H}_2)=n({\rm H_2O}^+)[n({\rm H}_2)k({\rm H_2O}^+|{\rm H}_2)+n_ek({\rm H_2O}^+|e^-)].
\label{eq_H2O+_steadystate}
\end{equation}
This can be re-arranged to produce the abundance ratio relation presented in \citet{gerin2010} and \citet{neufeld2010}:
\begin{equation}
\frac{n({\rm OH}^+)}{n({\rm H_2O}^+)}=\frac{k({\rm H_2O}^+|{\rm H}_2)}{k({\rm OH}^+|{\rm H}_2)}+\frac{2x_e}{f_{{\rm H}_2}}\frac{k({\rm H_2O}^+|e^-)}{k({\rm OH}^+|{\rm H}_2)},
\label{eq_OH+_over_H2O+}
\end{equation}
where 
$f_{\rm H_2}\equiv 2n({\rm H}_2)/n_{\rm H}$.  From this equation, the molecular hydrogen fraction, $f_{\rm H_2}$, can be determined using the abundances of OH$^+$, H$_2$O$^+$, and electrons, and the relevant reaction rate coefficients as
\begin{equation}
f_{\rm H_{2}}=\frac{2x_{e}k({\rm H_2O}^+|e^-)/k({\rm OH}^+|{\rm H}_2)}{N({\rm OH^{+}})/N({\rm H_{2}O^{+}})-k({\rm H_2O}^+|{\rm H}_2)/k({\rm OH}^+|{\rm H}_2)},
\label{eq_hydride_fH2}
\end{equation}
assuming constant densities and temperature in the region probed.  Given the OH$^+$ and H$_2$O$^+$ column densities found from our HIFI observations and the relevant rate coefficients from Table \ref{tbl_reactions} assuming $T=100$~K \citep[H~\textsc{i} spin temperature adopted by][]{godard2010}, we find a molecular hydrogen fraction of $f_{\rm H_2}=0.04\pm0.01$.\footnote{If $x_e=2x({\rm C}^+)$ in atomic regions as suggested by the \citet{hollenbach2012} models, $f_{\rm H_2}=0.08$.}  This is comparable to the low molecular fractions inferred toward W49N \citep{neufeld2010} and W31C \citep{gerin2010} from similar observations.  However, this is much lower than the cloud-averaged molecular fraction of $f_{\rm H_2}^N=2N({\rm H}_2)/(N({\rm H})+2N({\rm H}_2))=0.60$ found using the H and H$_2$ column densities reported in Table \ref{tbl_allmol}.  As such, OH$^+$ and H$_2$O$^+$ must  reside in the primarily atomic, outer layers of the observed cloud(s).

\subsection{Ionization Rate from OH$^+$ and H$_2$O$^+$}

Steady-state chemistry for OH$^+$ is given by
\begin{equation}
\epsilon\zeta_{\rm H}n({\rm H})=n({\rm OH}^+)[n({\rm H}_2)k({\rm OH}^+|{\rm H}_2)+n_ek({\rm OH}^+|e^-)],
\label{eq_OH+_steadystate}
\end{equation}
where $\zeta_{\rm H}$ is the cosmic-ray ionization rate of atomic hydrogen.  Here, the destruction of OH$^+$ is thought to include all important reactions, but not every H$^+$ formed by cosmic-ray ionization will eventually lead to OH$^+$.  To accommodate this fact, we introduce an efficiency factor, $\epsilon$, following \citet{neufeld2010}.  The quantity $\epsilon$ is defined as the ratio of the OH$^+$ formation rate to the cosmic-ray ionization rate of H.  Solving for the product $\epsilon\zeta_{\rm H}$ yields the equation
\begin{equation}
\epsilon\zeta_{\rm H}=\frac{N({\rm OH}^+)}{N({\rm H})}n_{\rm H}\left[\frac{f_{\rm H_2}}{2}k({\rm OH}^+|{\rm H}_2)+x_ek({\rm OH}^+|e^-) \right].
\label{eq_OH+_zeta}
\end{equation}
Because the OH$^+$ is thought to reside in the outskirts of the cloud, while the H$_3^+$ resides in the molecular interior, we use a simple pressure balance argument to estimate $n_{\rm H}$ in the atomic gas.  For a purely atomic exterior and a purely molecular interior, pressure balance requires the relation $T_{a}n_{a}=T_{m}n_{m}$, where the number density of collision partners in the atomic gas is equal to the hydrogen nucleon density in the atomic region ($n_{a}=n_{{\rm H},a}$), and the number density of collision partners in the molecular gas is equal to one-half of the hydrogen nucleon density in the molecular region ($n_{m}=n_{{\rm H},m}/2$).  Taking $T_a=100$~K, $T_m=70$~K, and $n_{{\rm H},m}=100$~cm$^{-3}$, we find $n_{{\rm H},a}=35$~cm$^{-3}$.  This value may be used for $n_{\rm H}$ in equation (\ref{eq_OH+_zeta}), along with the relevant column densities and rate coefficients, to obtain $\epsilon\zeta_{\rm H}=(0.21\pm0.11)\times10^{-16}$~s$^{-1}$.  Using the scaling between $\zeta_2$ and $\zeta_{\rm H}$ given by \citet{glassgold1974}, $1.5\zeta_2=2.3\zeta_{\rm H}$, and taking the value of $\zeta_2$ inferred from H$_3^+$ and $\epsilon\zeta_{\rm H}$ inferred from OH$^+$ and H$_2$O$^+$, we find an efficiency factor of $\epsilon=0.07\pm0.04$ for the production of OH$^+$ via cosmic-ray ionization of atomic hydrogen.  By combining equations (\ref{eqzeta2}), (\ref{eq_hydride_fH2}), and (\ref{eq_OH+_zeta}) the efficiency factor can be written as
\begin{multline}
\epsilon=\frac{2.3}{1.5}\frac{T_m}{2T_a}\frac{N({\rm OH}^+)}{N({\rm H})}\frac{N({\rm H}_2)}{N({\rm H}_3^+)}\frac{1}{k({\rm H}_3^+|e^-)}\times\\ 
\left[\frac{k({\rm H_2O}^+|e^-)}{N({\rm OH^{+}})/N({\rm H_{2}O^{+}})-k({\rm H_2O}^+|{\rm H}_2)/k({\rm OH}^+|{\rm H}_2)}+k({\rm OH}^+|e^-) \right],
\label{eq_epsilon}
\end{multline}
demonstrating that $\epsilon$ is independent of $x_e$ and $n_{\rm H}$.  Note, however, that if $x_e$ differs between diffuse atomic and diffuse molecular gas as suggested by \citet{hollenbach2012}, then a scaling factor corresponding to $x_{e,a}/x_{e,m}$ must be added to equation (\ref{eq_epsilon}).  As mentioned above, this effect may result in $x_{e,a}/x_{e,m}\sim2$, in which case $\epsilon$ would increase to 0.14.  Given the $T^{-0.5}$ dependence of the dissociative recombination rate coefficients of H$_3^+$, OH$^+$, and H$_2$O$^+$ the efficiency factor scales as $\epsilon\propto(T_a/T_m)^{-1.5}$ for different ratios between the temperatures in the atomic and molecular gas.

\section{DISCUSSION}

The efficiency factor we determine is much lower than that predicted by PDR models computed using the Meudon code \citep{lepetit2006,goicoechea2007}, where  $0.5\lesssim\epsilon\leq1.0$ \citep[see discussion in][]{neufeld2010}.  In those models, the chain of reactions leading from H$^+$ to OH$^+$ is broken by recombination of H$^+$ or O$^+$ with electrons, both of which decrease $\epsilon$.  Although recombination of O$^+$ is not important, recombination of H$^+$, while slow\footnote{$k({\rm H}^+|e^-)=3.5\times10^{-12}(T/300)^{-0.75}$~cm$^{3}$~s$^{-1}$; UDFA06}, can compete with the O$^+$ + H$_2$ reaction at low molecular fraction where the $\mathrm{O^+ + H\rightarrow O + H^+}$ back-reaction dominates the reaction with H$_2$ that forms OH$^+$.  As a result, the reaction network cycles between H$^+$ and O$^+$, sometimes forming OH$^+$ and sometimes forming H.  Because of this mechanism, at $f_{\rm H_2}=0.04$ the efficiency factor is about 0.5.

Another property of the Meudon models that can reduce $\epsilon$ is the inclusion of state-specific rate coefficients for the H$^+$ + O reaction, and the relative populations in the fine-structure levels of $(^3{\rm P}_J)$O.  The ground ($J=2$) level is likely the most populated state, and the H$^+$ + $(^3{\rm P}_2)$O reaction is dramatically slower at low ($\lesssim100$~K) temperatures than reactions involving $(^3{\rm P}_1)$O or $(^3{\rm P}_0)$O \citep{stancil1999}.  If charge transfer to O$^+$ is inhibited, then H$^+$ has more time to recombine with electrons, thus decreasing the efficiency at which OH$^+$ forms.


Still, recombination of H$^+$ alone cannot account for the small value of $\epsilon=0.07$ that we find.  In order to reach lower values of $\epsilon$ under diffuse cloud conditions, something other than electrons must be removing H$^+$ from the gas phase.  One possible mechanism for doing just this is the neutralization of H$^+$ on small grains or PAH's \citep{liszt2003}.  Recent modeling efforts \citep{hollenbach2012} that account for grain/PAH neutralization find values of $\epsilon\sim0.1$--0.3, much closer to our observationally-derived value.  We can apply these low efficiency factors to update ionization rates inferred in previous studies of OH$^+$ and H$_2$O$^+$.  Rescaling the lower limit on the ionization rate reported in \citet{gerin2010} to account for $\epsilon=0.07$, we find $\zeta_{\rm H}>2.6\times10^{-18}n({\rm H})$~s$^{-1}$ for the line of sight toward W31C.  Doing the same for the range of ionization rates reported in \citet{neufeld2010} toward W49N results in $8.6\times10^{-16}$~s$^{-1}\leq\zeta_{\rm H}\leq17\times10^{-16}$~s$^{-1}$.  These values are high compared to the distribution of ionization rates found using H$_3^+$ \citep[$\zeta_{\rm H}=(2.3_{-2.0}^{+3.4})\times10^{-16}$~s$^{-1}$, converted from the mean value of $\zeta_2$ in][]{indriolo2012}.  However, this scaling procedure is highly uncertain at present, and must be improved by determining $\epsilon$ in more cloud components.

It is also possible that our analysis underestimates $\epsilon$ given uncertainty in the rate coefficient for dissociative recombination of OH$^+$ with electrons.  As shown in Table \ref{tbl_reactions}, this process is 10 times slower than all of the other dissociative recombination reactions.  However, it has not been measured at temperatures relevant to diffuse interstellar clouds using vibrationally cold OH$^+$ molecules.  Under such conditions, it is possible that resonance structure in the low-energy collision cross section may increase the low-temperature rate of this reaction.  Indeed, some resonance structure has been observed \citep{amitay1996}, but that experiment did not determine the cross section on an absolute scale.  Future measurements of the OH$^+$ + $e^-$ dissociative recombination cross section using storage ring facilities are urgently needed.\footnote{If $k({\rm OH}^+|e^-)$ is 10 times larger than the value adopted in this study (see Table \ref{tbl_reactions}), then the value of $\epsilon$ required to bring the cosmic-ray ionization rates inferred from H$_3^+$ and OH$^+$ into agreement increases from 0.07 to 0.23}

Additionally, it would be advantageous to employ a chemical model that is more complete than the analytical expressions used herein.  Several species have been observed in the diffuse molecular cloud toward W51, all of which can be used in constraining the ambient physical conditions.  However, some species---e.g., H \citep{koo1997}, CH$^+$ \citep{falgarone2010}, OH$^+$, H$_2$O$^+$---are thought to reside in primarily atomic gas, while others---e.g., CH \citep{gerin2010ch}, HF \citep{sonnentrucker2010}, H$_2$O \citep{neufeld2002,sonnentrucker2010}, HCO$^+$, HCN \citep{godard2010}, CO---prefer molecular environments.  It seems apparent then, that even with similar velocity profiles, not all of the observed species are spatially co-located.  Instead, these atoms and molecules are likely probing different portions of a cloud complex, including diffuse atomic outer layers, and a diffuse molecular interior.  Any chemical model attempting to reproduce the observed abundances along this sight line must account for these effects.

Our analysis of OH$^+$ and H$_2$O$^+$ shows that both species must reside in gas of low molecular hydrogen fraction ($f_{\rm H_2}=0.04$) in order to explain the observed abundance ratio between the two species.  H$_3^+$, however, is not efficiently formed in gas with low molecular fractions where the H$_2^+$ + H reaction competes with the H$_2^+$ + H$_2$ reaction, meaning that H$_3^+$ must primarily reside in regions of higher molecular hydrogen fraction.  As such, it seems likely that there is little overlap in the gas probed by OH$^+$ and H$_2$O$^+$, and that probed by H$_3^+$ in diffuse molecular clouds (this changes for gas with a low electron fraction where the H$_3^+$ + O reaction becomes important).  

\section{SUMMARY}

We have made observations of H$_3^+$, OH$^+$, H$_2$O$^+$, and CO in a diffuse molecular cloud along closely spaced sight lines toward W51 IRS2.  The cosmic-ray ionization rate of H$_2$ inferred from the H$_3^+$ column density is $\zeta_2=(4.8\pm3.4)\times10^{-16}$~s$^{-1}$.   Observed OH$^+$ and H$_2$O$^+$ abundances yield an estimated molecular hydrogen fraction of $f_{\rm H_2}=0.04\pm0.01$ and a product of $\epsilon\zeta_{\rm H}=(0.21\pm0.11)\times10^{-16}$~s$^{-1}$ in the atomic outskirts of the cloud, where $\epsilon$ is defined following \citet{neufeld2010} as the ratio of the OH$^+$ production rate to the cosmic-ray ionization rate of H.  Combining both ionization rates, we find $\epsilon=0.07\pm0.04$, such that only 7\% of H$^+$ formed by cosmic-ray ionization goes on to eventually form OH$^+$.  A possible explanation for the low OH$^+$ formation efficiency is the neutralization of H$^+$ on small grains and PAHs, as suggested by \citet{hollenbach2012}.  Detailed chemical modeling that accounts for this process, the change from atomic to molecular gas with cloud depth, and all of the species observed in this diffuse molecular cloud, should provide further insight regarding the extent to which H$_3^+$ coexists with OH$^+$ and H$_2$O$^+$.

\mbox{}
The authors thank Takeshi Oka and Ben McCall for helpful discussion and suggestions, and for their part in the observations performed at UKIRT and Gemini South.  This research was performed, in part, through a JPL contract funded by the National Aeronautics and Space Administration.  M.G. acknowledges funding by CNES (Centre National d'Etudes Spatiales).  J.H.B. thanks the Swedish National Space Board for generous support.  J.R.G. acknowledges a Ram\'{o}n y Cajal research contract and thanks the Spanish MICINN for funding support through grants AYA2009-07304 and CSD2009-00038.  The United Kingdom Infrared Telescope is operated by the Joint Astronomy Centre on behalf of the Science and Technology Facilities Council of the U.K.  Based on observations obtained at the Gemini Observatory, which is operated by the Association of Universities for Research in Astronomy, Inc., under a cooperative agreement with the NSF on behalf of the Gemini partnership: the National Science Foundation (United States), the Science and Technology Facilities Council (United Kingdom), the National Research Council (Canada), CONICYT (Chile), the Australian Research Council (Australia), Minist\'{e}rio da Ci\^{e}ncia, Tecnologia e Inova\c{c}\~{a}o (Brazil) and Ministerio de Ciencia, Tecnolog\'{i}a e Innovaci\'{o}n Productiva (Argentina).  Gemini/Phoenix spectra were obtained through program GS-2010A-C-3.  This work is also based in part on observations obtained with the Phoenix infrared spectrograph, developed and operated by the National Optical Astronomy Observatory.  HIFI has been designed and built by a consortium of institutes and university departments from across Europe, Canada and the United States under the leadership of SRON Netherlands Institute for Space Research, Groningen, The Netherlands and with major contributions from Germany, France and the US. Consortium members are: Canada: CSA, U.Waterloo; France: CESR, LAB, LERMA, IRAM; Germany: KOSMA, MPIfR, MPS; Ireland, NUI Maynooth; Italy: ASI, IFSI-INAF, Osservatorio Astrofisico di Arcetri-INAF; Netherlands: SRON, TUD; Poland: CAMK, CBK; Spain: Observatorio Astronómico Nacional (IGN), Centro de Astrobiología (CSIC-INTA). Sweden: Chalmers University of Technology - MC2, RSS \& GARD; Onsala Space Observatory; Swedish National Space Board, Stockholm University - Stockholm Observatory; Switzerland: ETH Zurich, FHNW; USA: Caltech, JPL, NHSC.


\begin{thebibliography}{55}
\expandafter\ifx\csname natexlab\endcsname\relax\def\natexlab#1{#1}\fi

\bibitem[{{Amitay} {et~al.}(1996){Amitay}, {Zajfman}, {Forck}, {Heupel},
  {Grieser}, {Habs}, {Repnow}, {Schwalm}, {Wolf}, \& {Guberman}}]{amitay1996}
{Amitay}, Z., {Zajfman}, D., {Forck}, P., {et~al.} 1996, \pra, 53, 644

\bibitem[{{Bieging} {et~al.}(2010){Bieging}, {Peters}, \& {Kang}}]{bieging2010}
{Bieging}, J.~H., {Peters}, W.~L., \& {Kang}, M. 2010, \apjs, 191, 232

\bibitem[{{Carpenter} \& {Sanders}(1998)}]{carpenter1998}
{Carpenter}, J.~M., \& {Sanders}, D.~B. 1998, \aj, 116, 1856

\bibitem[{{de Graauw} {et~al.}(2010){de Graauw}, {Helmich}, {Phillips},
  {Stutzki}, {Caux}, {Whyborn}, {Dieleman}, {Roelfsema}, {Aarts}, {Assendorp},
  {Bachiller}, {Baechtold}, {Barcia}, {Beintema}, {Belitsky}, {Benz}, {Bieber},
  {Boogert}, {Borys}, {Bumble}, {Ca{\"i}s}, {Caris}, {Cerulli-Irelli},
  {Chattopadhyay}, {Cherednichenko}, {Ciechanowicz}, {Coeur-Joly}, {Comito},
  {Cros}, {de Jonge}, {de Lange}, {Delforges}, {Delorme}, {den Boggende},
  {Desbat}, {Diez-Gonz{\'a}lez}, {di Giorgio}, {Dubbeldam}, {Edwards},
  {Eggens}, {Erickson}, {Evers}, {Fich}, {Finn}, {Franke}, {Gaier}, {Gal},
  {Gao}, {Gallego}, {Gauffre}, {Gill}, {Glenz}, {Golstein}, {Goulooze},
  {Gunsing}, {G{\"u}sten}, {Hartogh}, {Hatch}, {Higgins}, {Honingh}, {Huisman},
  {Jackson}, {Jacobs}, {Jacobs}, {Jarchow}, {Javadi}, {Jellema}, {Justen},
  {Karpov}, {Kasemann}, {Kawamura}, {Keizer}, {Kester}, {Klapwijk}, {Klein},
  {Kollberg}, {Kooi}, {Kooiman}, {Kopf}, {Krause}, {Krieg}, {Kramer},
  {Kruizenga}, {Kuhn}, {Laauwen}, {Lai}, {Larsson}, {Leduc}, {Leinz}, {Lin},
  {Liseau}, {Liu}, {Loose}, {L{\'o}pez-Fernandez}, {Lord}, {Luinge}, {Marston},
  {Mart{\'{\i}}n-Pintado}, {Maestrini}, {Maiwald}, {McCoey}, {Mehdi}, {Megej},
  {Melchior}, {Meinsma}, {Merkel}, {Michalska}, {Monstein}, {Moratschke},
  {Morris}, {Muller}, {Murphy}, {Naber}, {Natale}, {Nowosielski}, {Nuzzolo},
  {Olberg}, {Olbrich}, {Orfei}, {Orleanski}, {Ossenkopf}, {Peacock}, {Pearson},
  {Peron}, {Phillip-May}, {Piazzo}, {Planesas}, {Rataj}, {Ravera}, {Risacher},
  {Salez}, {Samoska}, {Saraceno}, {Schieder}, {Schlecht}, {Schl{\"o}der},
  {Schm{\"u}lling}, {Schultz}, {Schuster}, {Siebertz}, {Smit}, {Szczerba},
  {Shipman}, {Steinmetz}, {Stern}, {Stokroos}, {Teipen}, {Teyssier}, {Tils},
  {Trappe}, {van Baaren}, {van Leeuwen}, {van de Stadt}, {Visser}, {Wildeman},
  {Wafelbakker}, {Ward}, {Wesselius}, {Wild}, {Wulff}, {Wunsch}, {Tielens},
  {Zaal}, {Zirath}, {Zmuidzinas}, \& {Zwart}}]{degraauw2010}
{de Graauw}, T., {Helmich}, F.~P., {Phillips}, T.~G., {et~al.} 2010, \aap, 518,
  L6

\bibitem[{{Falgarone} {et~al.}(2010){Falgarone}, {Godard}, {Cernicharo}, {de
  Luca}, {Gerin}, {Phillips}, {Black}, {Lis}, {Bell}, {Boulanger}, {Coutens},
  {Dartois}, {Encrenaz}, {Giesen}, {Goicoechea}, {Goldsmith}, {Gupta}, {Gry},
  {Hennebelle}, {Herbst}, {Hily-Blant}, {Joblin}, {Ka{\'z}mierczak},
  {Ko{\l}os}, {Kre{\l}owski}, {Martin-Pintado}, {Monje}, {Mookerjea},
  {Neufeld}, {Perault}, {Pearson}, {Persson}, {Plume}, {Salez}, {Schmidt},
  {Sonnentrucker}, {Stutzki}, {Teyssier}, {Vastel}, {Yu}, {Menten}, {Geballe},
  {Schlemmer}, {Shipman}, {Tielens}, {Philipp}, {Cros}, {Zmuidzinas},
  {Samoska}, {Klein}, {Lorenzani}, {Szczerba}, {P{\'e}ron}, {Cais}, {Gaufre},
  {Cros}, {Ravera}, {Morris}, {Lord}, \& {Planesas}}]{falgarone2010}
{Falgarone}, E., {Godard}, B., {Cernicharo}, J., {et~al.} 2010, \aap, 521, L15

\bibitem[{{Geballe} \& {Oka}(2010)}]{geballe2010}
{Geballe}, T.~R., \& {Oka}, T. 2010, \apjl, 709, L70

\bibitem[{{Gerin}(2012)}]{gerin2012}
{Gerin}, M. 2012, Philos. Trans. R. Soc. London A, in press

\bibitem[{{Gerin} {et~al.}(2011){Gerin}, {Ka{\'z}mierczak}, {Jastrzebska},
  {Falgarone}, {Hily-Blant}, {Godard}, \& {de Luca}}]{gerin2011}
{Gerin}, M., {Ka{\'z}mierczak}, M., {Jastrzebska}, M., {et~al.} 2011, \aap,
  525, A116

\bibitem[{{Gerin} {et~al.}(2010{\natexlab{a}}){Gerin}, {de Luca}, {Goicoechea},
  {Herbst}, {Falgarone}, {Godard}, {Bell}, {Coutens}, {Ka{\'z}mierczak},
  {Sonnentrucker}, {Black}, {Neufeld}, {Phillips}, {Pearson}, {Rimmer},
  {Hassel}, {Lis}, {Vastel}, {Boulanger}, {Cernicharo}, {Dartois}, {Encrenaz},
  {Giesen}, {Goldsmith}, {Gupta}, {Gry}, {Hennebelle}, {Hily-Blant}, {Joblin},
  {Ko{\l}os}, {Kre{\l}owski}, {Mart{\'{\i}}n-Pintado}, {Monje}, {Mookerjea},
  {Perault}, {Persson}, {Plume}, {Salez}, {Schmidt}, {Stutzki}, {Teyssier},
  {Yu}, {Contursi}, {Menten}, {Geballe}, {Schlemmer}, {Morris}, {Hatch},
  {Imram}, {Ward}, {Caux}, {G{\"u}sten}, {Klein}, {Roelfsema}, {Dieleman},
  {Schieder}, {Honingh}, \& {Zmuidzinas}}]{gerin2010ch}
{Gerin}, M., {de Luca}, M., {Goicoechea}, J.~R., {et~al.} 2010{\natexlab{a}},
  \aap, 521, L16

\bibitem[{{Gerin} {et~al.}(2010{\natexlab{b}}){Gerin}, {de Luca}, {Black},
  {Goicoechea}, {Herbst}, {Neufeld}, {Falgarone}, {Godard}, {Pearson}, {Lis},
  {Phillips}, {Bell}, {Sonnentrucker}, {Boulanger}, {Cernicharo}, {Coutens},
  {Dartois}, {Encrenaz}, {Giesen}, {Goldsmith}, {Gupta}, {Gry}, {Hennebelle},
  {Hily-Blant}, {Joblin}, {Kazmierczak}, {Kolos}, {Krelowski},
  {Martin-Pintado}, {Monje}, {Mookerjea}, {Perault}, {Persson}, {Plume},
  {Rimmer}, {Salez}, {Schmidt}, {Stutzki}, {Teyssier}, {Vastel}, {Yu},
  {Contursi}, {Menten}, {Geballe}, {Schlemmer}, {Shipman}, {Tielens},
  {Philipp-May}, {Cros}, {Zmuidzinas}, {Samoska}, {Klein}, \&
  {Lorenzani}}]{gerin2010}
{Gerin}, M., {de Luca}, M., {Black}, J., {et~al.} 2010{\natexlab{b}}, \aap,
  518, L110

\bibitem[{{Glassgold} \& {Langer}(1974)}]{glassgold1974}
{Glassgold}, A.~E., \& {Langer}, W.~D. 1974, \apj, 193, 73

\bibitem[{{Godard} {et~al.}(2010){Godard}, {Falgarone}, {Gerin}, {Hily-Blant},
  \& {de Luca}}]{godard2010}
{Godard}, B., {Falgarone}, E., {Gerin}, M., {Hily-Blant}, P., \& {de Luca}, M.
  2010, \aap, 520, A20

\bibitem[{{Goicoechea} \& {Le Bourlot}(2007)}]{goicoechea2007}
{Goicoechea}, J.~R., \& {Le Bourlot}, J. 2007, \aap, 467, 1

\bibitem[{{Goto} {et~al.}(2002){Goto}, {McCall}, {Geballe}, {Usuda},
  {Kobayashi}, {Terada}, \& {Oka}}]{goto2002}
{Goto}, M., {McCall}, B.~J., {Geballe}, T.~R., {et~al.} 2002, \pasj, 54, 951

\bibitem[{{Goto} {et~al.}(2011){Goto}, {Usuda}, {Geballe}, {Indriolo},
  {McCall}, {Henning}, \& {Oka}}]{goto2011}
{Goto}, M., {Usuda}, T., {Geballe}, T.~R., {et~al.} 2011, PASJ, 63, L13

\bibitem[{{Goto} {et~al.}(2008){Goto}, {Usuda}, {Nagata}, {Geballe}, {McCall},
  {Indriolo}, {Suto}, {Henning}, {Morong}, \& {Oka}}]{goto2008}
{Goto}, M., {Usuda}, T., {Nagata}, T., {et~al.} 2008, \apj, 688, 306

\bibitem[{{Hinkle} {et~al.}(2003){Hinkle}, {Blum}, {Joyce}, {Sharp}, {Ridgway},
  {Bouchet}, {van der Bliek}, {Najita}, \& {Winge}}]{hinkle2003}
{Hinkle}, K.~H., {Blum}, R.~D., {Joyce}, R.~R., {et~al.} 2003, \procspie, 4834,
  353

\bibitem[{{Hollenbach} {et~al.}(2012){Hollenbach}, {Kaufman}, {Neufeld},
  {Wolfire}, \& {Goicoechea}}]{hollenbach2012}
{Hollenbach}, D., {Kaufman}, M.~J., {Neufeld}, D.~A., {Wolfire}, M.~G., \&
  {Goicoechea}, J.~R. 2012, ApJ in press (arXiv:1205.6446v1)

\bibitem[{{Indriolo}(2011)}]{indriolo2011}
{Indriolo}, N. 2011, PhD thesis, University of Illinois

\bibitem[{{Indriolo} {et~al.}(2007){Indriolo}, {Geballe}, {Oka}, \&
  {McCall}}]{indriolo2007}
{Indriolo}, N., {Geballe}, T.~R., {Oka}, T., \& {McCall}, B.~J. 2007, \apj,
  671, 1736

\bibitem[{{Indriolo} \& {McCall}(2012)}]{indriolo2012}
{Indriolo}, N., \& {McCall}, B.~J. 2012, \apj, 745, 91

\bibitem[{{Jensen} {et~al.}(2000){Jensen}, {Bilodeau}, {Safvan}, {Seiersen},
  {Andersen}, {Pedersen}, \& {Heber}}]{jensen2000}
{Jensen}, M.~J., {Bilodeau}, R.~C., {Safvan}, C.~P., {et~al.} 2000, \apj, 543,
  764

\bibitem[{{Jones} {et~al.}(1981){Jones}, {Birkinshaw}, \& {Twiddy}}]{jones1981}
{Jones}, J.~D.~C., {Birkinshaw}, K., \& {Twiddy}, N.~D. 1981, Chemical Physics
  Letters, 77, 484

\bibitem[{{Kang} {et~al.}(2010){Kang}, {Bieging}, {Kulesa}, {Lee}, {Choi}, \&
  {Peters}}]{kang2010}
{Kang}, M., {Bieging}, J.~H., {Kulesa}, C.~A., {et~al.} 2010, \apjs, 190, 58

\bibitem[{{Koo}(1997)}]{koo1997}
{Koo}, B.-C. 1997, \apjs, 108, 489

\bibitem[{{Kreckel} {et~al.}(2005){Kreckel}, {Motsch}, {Mikosch},
  {Glos{\'{\i}}k}, {Pla{\v s}il}, {Altevogt}, {Andrianarijaona}, {Buhr},
  {Hoffmann}, {Lammich}, {Lestinsky}, {Nevo}, {Novotny}, {Orlov}, {Pedersen},
  {Sprenger}, {Terekhov}, {Toker}, {Wester}, {Gerlich}, {Schwalm}, {Wolf}, \&
  {Zajfman}}]{kreckel2005}
{Kreckel}, H., {Motsch}, M., {Mikosch}, J., {et~al.} 2005, Physical Review
  Letters, 95, 263201

\bibitem[{{Kreckel} {et~al.}(2010){Kreckel}, {Novotn{\'y}}, {Crabtree}, {Buhr},
  {Petrignani}, {Tom}, {Thomas}, {Berg}, {Bing}, {Grieser}, {Krantz},
  {Lestinsky}, {Mendes}, {Nordhorn}, {Repnow}, {St{\"u}tzel}, {Wolf}, \&
  {McCall}}]{kreckel2010}
{Kreckel}, H., {Novotn{\'y}}, O., {Crabtree}, K.~N., {et~al.} 2010, \pra, 82,
  042715

\bibitem[{{Le Petit} {et~al.}(2006){Le Petit}, {Nehm{\'e}}, {Le Bourlot}, \&
  {Roueff}}]{lepetit2006}
{Le Petit}, F., {Nehm{\'e}}, C., {Le Bourlot}, J., \& {Roueff}, E. 2006, \apjs,
  164, 506

\bibitem[{{Liszt}(2003)}]{liszt2003}
{Liszt}, H. 2003, \aap, 398, 621

\bibitem[{{McCall}(2001)}]{mccall2001}
{McCall}, B.~J. 2001, PhD thesis, The University of Chicago

\bibitem[{{McCall} {et~al.}(2004){McCall}, {Huneycutt}, {Saykally}, {Djuric},
  {Dunn}, {Semaniak}, {Novotny}, {Al-Khalili}, {Ehlerding}, {Hellberg},
  {Kalhori}, {Neau}, {Thomas}, {Paal}, {{\"O}sterdahl}, \&
  {Larsson}}]{mccall2004}
{McCall}, B.~J., {Huneycutt}, A.~J., {Saykally}, R.~J., {et~al.} 2004, Phys.
  Rev. A, 70, 052716

\bibitem[{{Mehringer}(1994)}]{mehringer1994}
{Mehringer}, D.~M. 1994, \apjs, 91, 713

\bibitem[{{Mitchell}(1990)}]{mitchell1990}
{Mitchell}, J.~B.~A. 1990, \physrep, 186, 215

\bibitem[{{Mountain} {et~al.}(1990){Mountain}, {Robertson}, {Lee}, \&
  {Wade}}]{mountain1990}
{Mountain}, C.~M., {Robertson}, D.~J., {Lee}, T.~J., \& {Wade}, R. 1990,
  \procspie, 1235, 25

\bibitem[{{M{\"u}ller} {et~al.}(2005){M{\"u}ller}, {Schl{\"o}der}, {Stutzki},
  \& {Winnewisser}}]{muller2005}
{M{\"u}ller}, H.~S.~P., {Schl{\"o}der}, F., {Stutzki}, J., \& {Winnewisser}, G.
  2005, Journal of Molecular Structure, 742, 215

\bibitem[{{M{\"u}rtz} {et~al.}(1998){M{\"u}rtz}, {Zink}, {Evenson}, \&
  {Brown}}]{murtz1998}
{M{\"u}rtz}, P., {Zink}, L.~R., {Evenson}, K.~M., \& {Brown}, J.~M. 1998, \jcp,
  109, 9744

\bibitem[{{Neufeld} {et~al.}(2002){Neufeld}, {Kaufman}, {Goldsmith},
  {Hollenbach}, \& {Plume}}]{neufeld2002}
{Neufeld}, D.~A., {Kaufman}, M.~J., {Goldsmith}, P.~F., {Hollenbach}, D.~J., \&
  {Plume}, R. 2002, \apj, 580, 278

\bibitem[{{Neufeld} {et~al.}(1995){Neufeld}, {Lepp}, \&
  {Melnick}}]{neufeld1995}
{Neufeld}, D.~A., {Lepp}, S., \& {Melnick}, G.~J. 1995, \apjs, 100, 132

\bibitem[{{Neufeld} {et~al.}(2010){Neufeld}, {Goicoechea}, {Sonnentrucker},
  {Black}, {Pearson}, {Yu}, {Phillips}, {Lis}, {de Luca}, {Herbst}, {Rimmer},
  {Gerin}, {Bell}, {Boulanger}, {Cernicharo}, {Coutens}, {Dartois},
  {Kazmierczak}, {Encrenaz}, {Falgarone}, {Geballe}, {Giesen}, {Godard},
  {Goldsmith}, {Gry}, {Gupta}, {Hennebelle}, {Hily-Blant}, {Joblin},
  {Ko{\l}os}, {Kre{\l}owski}, {Mart{\'{\i}}n-Pintado}, {Menten}, {Monje},
  {Mookerjea}, {Perault}, {Persson}, {Plume}, {Salez}, {Schlemmer}, {Schmidt},
  {Stutzki}, {Teyssier}, {Vastel}, {Cros}, {Klein}, {Lorenzani}, {Philipp},
  {Samoska}, {Shipman}, {Tielens}, {Szczerba}, \& {Zmuidzinas}}]{neufeld2010}
{Neufeld}, D.~A., {Goicoechea}, J.~R., {Sonnentrucker}, P., {et~al.} 2010,
  \aap, 521, L10

\bibitem[{{Oka} {et~al.}(2005){Oka}, {Geballe}, {Goto}, {Usuda}, \&
  {McCall}}]{oka2005}
{Oka}, T., {Geballe}, T.~R., {Goto}, M., {Usuda}, T., \& {McCall}, B.~J. 2005,
  \apj, 632, 882

\bibitem[{{Okumura} {et~al.}(2001){Okumura}, {Miyawaki}, {Sorai}, {Yamashita},
  \& {Hasegawa}}]{okumura2001}
{Okumura}, S.-I., {Miyawaki}, R., {Sorai}, K., {Yamashita}, T., \& {Hasegawa},
  T. 2001, \pasj, 53, 793

\bibitem[{{Pilbratt} {et~al.}(2010){Pilbratt}, {Riedinger}, {Passvogel},
  {Crone}, {Doyle}, {Gageur}, {Heras}, {Jewell}, {Metcalfe}, {Ott}, \&
  {Schmidt}}]{pilbratt2010}
{Pilbratt}, G.~L., {Riedinger}, J.~R., {Passvogel}, T., {et~al.} 2010, \aap,
  518, L1

\bibitem[{{Rachford} {et~al.}(2002){Rachford}, {Snow}, {Tumlinson}, {Shull},
  {Blair}, {Ferlet}, {Friedman}, {Gry}, {Jenkins}, {Morton}, {Savage},
  {Sonnentrucker}, {Vidal-Madjar}, {Welty}, \& {York}}]{rachford2002}
{Rachford}, B.~L., {Snow}, T.~P., {Tumlinson}, J., {et~al.} 2002, \apj, 577,
  221

\bibitem[{{Rachford} {et~al.}(2009){Rachford}, {Snow}, {Destree}, {Ross},
  {Ferlet}, {Friedman}, {Gry}, {Jenkins}, {Morton}, {Savage}, {Shull},
  {Sonnentrucker}, {Tumlinson}, {Vidal-Madjar}, {Welty}, \&
  {York}}]{rachford2009}
{Rachford}, B.~L., {Snow}, T.~P., {Destree}, J.~D., {et~al.} 2009, \apjs, 180,
  125

\bibitem[{Rakshit \& Warneck(1980)}]{rakshit1980}
Rakshit, A.~B., \& Warneck, P. 1980, J. Chem. Soc.{,} Faraday Trans. 2, 76,
  1084

\bibitem[{Rosen {et~al.}(2000)Rosen, Derkatch, Semaniak, Neau, Al-Khalili,
  Le~Padellec, Vikor, Thomas, Danared, af~Ugglas, \& Larsson}]{rosen2000}
Rosen, S., Derkatch, A., Semaniak, J., {et~al.} 2000, Faraday Discuss., 115,
  295

\bibitem[{{Savage} {et~al.}(1977){Savage}, {Bohlin}, {Drake}, \&
  {Budich}}]{savage1977}
{Savage}, B.~D., {Bohlin}, R.~C., {Drake}, J.~F., \& {Budich}, W. 1977, \apj,
  216, 291

\bibitem[{{Sheffer} {et~al.}(2008){Sheffer}, {Rogers}, {Federman}, {Abel},
  {Gredel}, {Lambert}, \& {Shaw}}]{sheffer2008}
{Sheffer}, Y., {Rogers}, M., {Federman}, S.~R., {et~al.} 2008, \apj, 687, 1075

\bibitem[{{Smith} {et~al.}(1978){Smith}, {Adams}, \& {Miller}}]{smith1978}
{Smith}, D., {Adams}, N.~G., \& {Miller}, T.~M. 1978, \jcp, 69, 308

\bibitem[{{Sonnentrucker} {et~al.}(2010){Sonnentrucker}, {Neufeld}, {Phillips},
  {Gerin}, {Lis}, {de Luca}, {Goicoechea}, {Black}, {Bell}, {Boulanger},
  {Cernicharo}, {Coutens}, {Dartois}, {Ka{\'z}mierczak}, {Encrenaz},
  {Falgarone}, {Geballe}, {Giesen}, {Godard}, {Goldsmith}, {Gry}, {Gupta},
  {Hennebelle}, {Herbst}, {Hily-Blant}, {Joblin}, {Ko{\l}os}, {Kre{\l}owski},
  {Mart{\'{\i}}n-Pintado}, {Menten}, {Monje}, {Mookerjea}, {Pearson},
  {Perault}, {Persson}, {Plume}, {Salez}, {Schlemmer}, {Schmidt}, {Stutzki},
  {Teyssier}, {Vastel}, {Yu}, {Caux}, {G{\"u}sten}, {Hatch}, {Klein}, {Mehdi},
  {Morris}, \& {Ward}}]{sonnentrucker2010}
{Sonnentrucker}, P., {Neufeld}, D.~A., {Phillips}, T.~G., {et~al.} 2010, \aap,
  521, L12

\bibitem[{{Stancil} {et~al.}(1999){Stancil}, {Schultz}, {Kimura}, {Gu},
  {Hirsch}, \& {Buenker}}]{stancil1999}
{Stancil}, P.~C., {Schultz}, D.~R., {Kimura}, M., {et~al.} 1999, A\&AS, 140,
  225

\bibitem[{{Theard} \& {Huntress}(1974)}]{theard1974}
{Theard}, L.~P., \& {Huntress}, W.~T. 1974, J. Chem. Phys., 60, 2840

\bibitem[{{Welsh} {et~al.}(2010){Welsh}, {Lallement}, {Vergely}, \&
  {Raimond}}]{welsh2010}
{Welsh}, B.~Y., {Lallement}, R., {Vergely}, J.-L., \& {Raimond}, S. 2010, \aap,
  510, A54

\bibitem[{{Woodall} {et~al.}(2007){Woodall}, {Ag{\'u}ndez}, {Markwick-Kemper},
  \& {Millar}}]{woodall2007}
{Woodall}, J., {Ag{\'u}ndez}, M., {Markwick-Kemper}, A.~J., \& {Millar}, T.~J.
  2007, \aap, 466, 1197

\bibitem[{{Wyrowski} {et~al.}(2010){Wyrowski}, {van der Tak}, {Herpin},
  {Baudry}, {Bontemps}, {Chavarria}, {Frieswijk}, {Jacq}, {Marseille},
  {Shipman}, {van Dishoeck}, {Benz}, {Caselli}, {Hogerheijde}, {Johnstone},
  {Liseau}, {Bachiller}, {Benedettini}, {Bergin}, {Bjerkeli}, {Blake},
  {Braine}, {Bruderer}, {Cernicharo}, {Codella}, {Daniel}, {di Giorgio},
  {Dominik}, {Doty}, {Encrenaz}, {Fich}, {Fuente}, {Giannini}, {Goicoechea},
  {de Graauw}, {Helmich}, {Herczeg}, {J{\o}rgensen}, {Kristensen}, {Larsson},
  {Lis}, {McCoey}, {Melnick}, {Nisini}, {Olberg}, {Parise}, {Pearson}, {Plume},
  {Risacher}, {Santiago}, {Saraceno}, {Tafalla}, {van Kempen}, {Visser},
  {Wampfler}, {Y{\i}ld{\i}z}, {Black}, {Falgarone}, {Gerin}, {Roelfsema},
  {Dieleman}, {Beintema}, {de Jonge}, {Whyborn}, {Stutzki}, \&
  {Ossenkopf}}]{wyrowski2010}
{Wyrowski}, F., {van der Tak}, F., {Herpin}, F., {et~al.} 2010, \aap, 521, L34

\end{thebibliography}



\clearpage
\begin{figure}
\epsscale{0.5}
\plotone{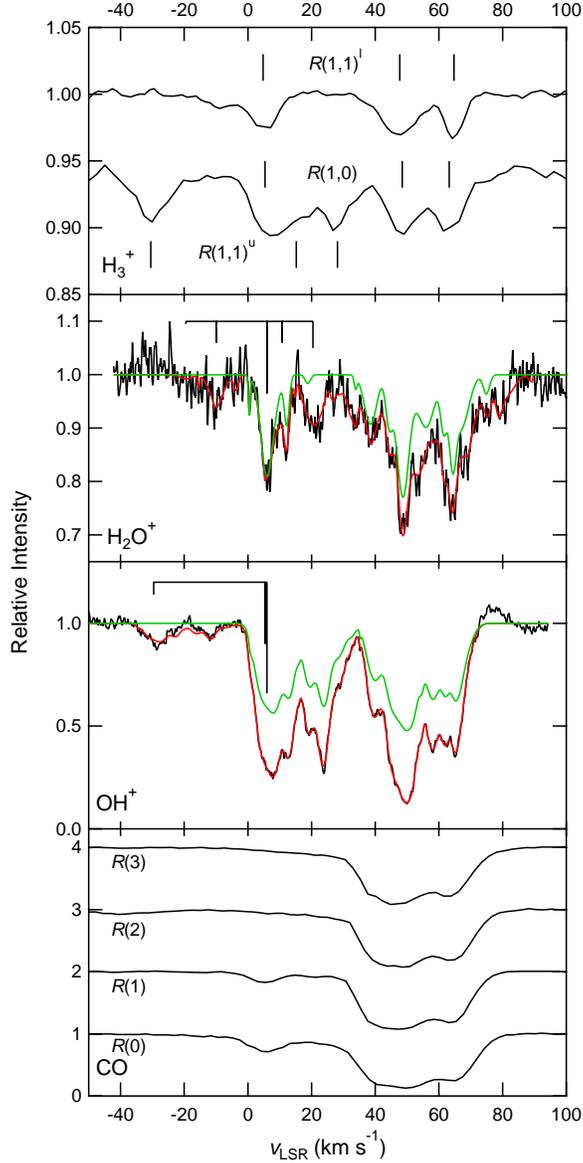}
\caption{Spectra toward W51 in velocity space.  {\bf First panel:} H$_3^+$ spectra showing the $R(1,1)^l$ line (top spectrum) and the $R(1,1)^u$ and $R(1,0)$ lines (bottom spectrum).  Vertical lines mark the absorption features at about 5~km~s$^{-1}$, 50~km~s$^{-1}$, and 65~km~s$^{-1}$.  In the bottom spectrum the $R(1,0)$ line is set to zero velocity, such that the $R(1,1)^u$ features are shifted $-35.4$~km~s$^{-1}$ from LSR velocity.  {\bf Second panel:} Spectrum showing the $N_{K_{a}K_{c}}=1_{11}-0_{00}$, $J=3/2-1/2$ transition of {\it ortho}-H$_2$O$^+$.  A stick diagram centered at 6~km~s$^{-1}$ shows the hyperfine structure of the observed transition.  The black curve is the observed spectrum, the red curve is the fit to that spectrum, and the green curve shows only the strongest hyperfine component portion of the fit.  {\bf Third panel:} Spectrum showing the $N=1-0$, $J=2-1$ transition of OH$^+$.  Colors are the same as for H$_2$O$^+$.  {\bf Fourth panel:}  Spectra of the $v$=1--0 band of $^{12}$CO showing (from top to bottom) the $R(3)$, $R(2)$, $R(1)$, and $R(0)$ transitions.}
\label{fig_spectra}

\end{figure}


\clearpage
\small
\begin{longtable}{llc}
\caption{Reaction Network}\\
\hline
\hline
              & Rate Coefficient &                       \\
Reaction & (cm$^3$~s$^{-1}$) & Reference \\
\hline
\endhead
\hline
\multicolumn{3}{p{6.0in}}{{\bf Notes:} Both $\zeta_2$ and $\zeta_{\rm H}$ have units of s$^{-1}$.  Rate coefficients are those used in UDFA06 (www.udfa.net).  All temperature dependent coefficients are in terms of the gas kinetic temperature, $T$, except for $k({\rm H}_3^+|e^-)$, which is in terms of the electron temperature, $T_e$.  However, it is expected that electrons and H$_2$ should be thermalized to the gas kinetic temperature via collisions, so we make no distinction between $T_e$ and $T$ in our calculations.\newline
$^{\dagger}$This expression includes contributions from the state-specific reactions H$^+$ + ($^3{\rm P}_ J$)O, where $J=0$, 1, and 2, and makes assumptions regarding the relative populations in these fine-structure levels.  Rate coefficients for the state-specific reactions were computed by \citet{stancil1999}.\newline
{\bf References:} (1) \citet{theard1974}; (2) \citet{mccall2004}; (3) \citet{woodall2007}; (4) \citet{stancil1999}; (5) \citet{smith1978}; (6) \citet{jones1981}; (7) \citet{mitchell1990}; (8) \citet{rakshit1980}; (9) \citet{rosen2000}; (10) \citet{jensen2000}
}
\endlastfoot
${\rm H}_2 + {\rm CR}\rightarrow {\rm H}_2^+ + e^- + {\rm CR}'$    & $\zeta_2$ &  \\
${\rm H}_2^+ + {\rm H}_2\rightarrow {\rm H}_3^+ + {\rm H}$          & $k({\rm H}_2^+|{\rm H}_2)=2.08\times10^{-9}$  & 1 \\
${\rm H}_3^+ + e^-\rightarrow {\rm H}_2 + {\rm H~or~H + H + H}$ & $k({\rm H}_3^+|e^-)=-1.3\times10^{-8}+1.27\times10^{-6}T_e^{-0.48}$ & 2 \\
\hline
${\rm H} + {\rm CR} \rightarrow {\rm H}^+ + e^- + {\rm CR'}$ & $\zeta_{\rm H}$ & \\
${\rm H}^+ + {\rm O} \rightarrow {\rm O}^+ + {\rm H}$ & $k({\rm H}^+|{\rm O})=7.31\times10^{-10}(T/300)^{0.23}\exp(-225.9/T)^{\dagger}$ & 3,4 \\
$\mathrm{O^+ + H_2 \rightarrow OH^+ + H}$ & $k({\rm O}^+|{\rm H}_2)=1.7\times10^{-9}$ & 5 \\
$\mathrm{OH^{+} + H_{2} \rightarrow H_{2}O^{+} + H}$ & $k({\rm OH}^+|{\rm H}_2)=1.01\times10^{-9}$ & 6 \\
$\mathrm{OH^{+}} + e^{-} \rightarrow \mathrm{products}$ & $k({\rm OH}^+|e^-)=3.75\times10^{-8}(T/300)^{-0.5}$ & 7 \\
$\mathrm{H_{2}O^{+} + H_{2} \rightarrow H_{3}O^{+} + H}$ & $k({\rm H_2O}^+|{\rm H}_2)=6.4\times10^{-10}$ & 8 \\
$\mathrm{H_{2}O^{+}} + e^{-} \rightarrow \mathrm{products}$ & $k({\rm H_2O}^+|e^-)=4.3\times10^{-7}(T/300)^{-0.5}$ & 9 \\
$\mathrm{H_{3}O^{+}} + e^{-} \rightarrow \mathrm{products}$ & $k({\rm H_3O}^+|e^-)=4.3\times10^{-7}(T/300)^{-0.5}$ & 10 
\label{tbl_reactions}
\end{longtable}
\normalsize


\clearpage
\small
\begin{longtable}{clcccccc}
\caption{IR Absorption Line Parameters}\\
\hline
\hline
               &                 & $v_{\rm LSR}$  & FWHM              & $W_{\lambda}$          & $\sigma(W_{\lambda})$ & $N(J,K)$                         &  $\sigma(N)$ \\
Molecule & Transition & (km~s$^{-1}$) & (km~s$^{-1}$) & ($10^{-6}$~$\mu$m) & ($10^{-6}$~$\mu$m)   & ($10^{13}$~cm$^{-2}$) & ($10^{13}$~cm$^{-2}$) \\
\hline
\endhead
\hline
\multicolumn{8}{p{6.0in}}{{\bf Notes:} Columns 3 and 4 are the line-center velocity and velocity full-width at half-maximum (including instrumental broadening effects) found by a Gaussian fit to the absorption feature.  Columns 5 and 6 are the equivalent width, $W_\lambda$, and its $1\sigma$ uncertainty.  Columns 7 and 8 are the column density in the lower state, $N$, and its $1\sigma$ uncertainty.  The $R(1,0)$ line from the 5~km~s$^{-1}$ cloud and the $R(1,1)^u$ line from the 50~km~s$^{-1}$ cloud are severely blended.  Note the difference in units for the equivalent widths and column densities of H$_3^+$ versus CO.  In calculating $\sigma(W_{\lambda})$ for the $R(2)$ and $R(3)$ lines a FWHM of 10~km~s$^{-1}$ is adopted.  Values reported for the $R(2)$ and $R(3)$ lines of CO are $1\sigma$ upper limits on the equivalent width and column density in the 5~km~s$^{-1}$ component.  CO column densities are calculated assuming optically thin conditions, and are likely lower limits.  For comparison, a curve-of-growth analysis with $b=2$~km~s$^{-1}$ results in $N(0)=1.97\times10^{15}$~cm$^{-2}$ and $N(1)=1.49\times10^{15}$~cm$^{-2}$.
}
\endlastfoot
H$_3^+$ & $R(1,1)^u$ &   4.9 & 11.5 & 4.98 & 0.64 & 20.7 & 2.65 \\
H$_3^+$ & $R(1,0)$     &   5.4 & 10.0 & 5.01 & 0.69 & 12.7 & 1.75 \\
H$_3^+$ & $R(1,1)^l$  &   4.8 & 10.1 & 3.49 & 0.14 & 16.0 & 0.64 \\
H$_3^+$ & $R(1,1)^u$ & 50.6 & 12.8 & 5.52 & 1.76 & 22.9 & 7.30 \\
H$_3^+$ & $R(1,0)$     & 48.5 & 11.5 & 6.51 & 1.69 & 16.5 & 4.28 \\
H$_3^+$ & $R(1,1)^l$  & 47.7 & 13.9 & 5.57 & 0.17 & 25.6 & 0.78 \\
H$_3^+$ & $R(1,1)^u$ & 63.5 & 10.0 & 4.82 & 1.33 & 20.0 & 5.52 \\
H$_3^+$ & $R(1,0)$     & 63.2 & 11.0 & 5.93 & 0.51 & 15.0 & 1.29 \\
H$_3^+$ & $R(1,1)^l$  & 64.7 &   6.7 & 2.92 & 0.12 & 13.4 & 0.55 \\
\hline
                &                 & $v_{\rm LSR}$  & FWHM              & $W_{\lambda}$          & $\sigma(W_{\lambda})$ & $N(J)_{\rm thin}$                         &  $\sigma(N_{\rm thin})$ \\
Molecule & Transition & (km~s$^{-1}$) & (km~s$^{-1}$) & ($10^{-5}$~$\mu$m) & ($10^{-5}$~$\mu$m)   & ($10^{15}$~cm$^{-2}$) & ($10^{15}$~cm$^{-2}$) \\
\hline
CO & $R(0)$ & 4.9 & 10.5 & 3.33 & 0.17 & 1.52 & 0.08 \\
CO & $R(1)$ & 4.7 &   8.7 & 1.90 & 0.28 & 1.30 & 0.19 \\
CO & $R(2)$ & ...   & ...     & ...     & 0.09 & ...     & 0.07 \\
CO & $R(3)$ & ...   & ...     & ...     & 0.05 & ...     & 0.04
\label{tbl_IRabsorption}
\end{longtable}
\normalsize

\clearpage
\begin{longtable}{cccccc}
\caption{Results from OH$^+$ and H$_2$O$^+$ Spectra} \\
$v_{\rm LSR}$  & $N({\rm OH}^+)$ & $\sigma[N({\rm OH}^+)]$ & $N(o$-H$_2$O$^+)$ & $\sigma[N(o$-H$_2$O$^+)]$ & \\
(km~s$^{-1}$) & ($10^{12}$~cm$^{-2}$) & ($10^{12}$~cm$^{-2}$) & ($10^{12}$~cm$^{-2}$) & ($10^{12}$~cm$^{-2}$) & $f_{\rm H_2}$ \\
\hline
\hline
\endhead
\hline
\multicolumn{6}{p{6.0in}}{{\bf Notes:} Velocity intervals were chosen to roughly correspond to the larger absorption components in the fit to the OH$^+$ and H$_2$O$^+$ spectra.  $f_{\rm H_2}$ is calculated using equation (\ref{eq_hydride_fH2}) and assuming an {\it ortho}-to-{\it para} ratio of 3 for H$_2$O$^+$.  In the 21--33~km~s$^{-1}$ component the $3\sigma$ uncertainty in the H$_2$O$^+$ column density is used to determine the upper limit on $f_{\rm H_2}$.  Observations of {\it ortho}-H$_2$O$^+$ toward W51 performed by the WISH (Water In Star-forming regions with {\it Herschel}) key program show absorption at 22.5~km~s$^{-1}$ at a position 60\arcsec\ away from the PRISMAS pointing (17\arcsec\ from the IR pointing).  The inferred column density is $N(o$-H$_2$O$^+)=0.38\times10^{12}$~cm$^{-2}$ \citep{wyrowski2010}---consistent with our upper limit---and, when taken with our OH$^+$ column density, gives a molecular hydrogen fraction of 0.02 in that component.  It should also be noted that \citet{wyrowski2010} find a column density of $N(o$-H$_2$O$^+)=4.5\times10^{12}$~cm$^{-2}$ in a component at 6~km~s$^{-1}$, in excellent agreement with our findings, validating our assumption that the slightly different sight lines targeted by the THz and IR observations should indeed probe similar material in the nearby diffuse cloud.
}
\endlastfoot
0--11   & 29.7 & 1.30 & 4.57 & 0.72 & 0.04 \\
11--17 & 11.7 & 0.55 & 0.77 & 0.31 & 0.02 \\
17--21 & 7.93 & 0.36 & 0.15 & 0.15 & 0.01 \\
21--33 & 18.7 & 1.00 &     ... & 0.64 &  $<0.03$   \\
33--42 & 8.94 & 0.56 & 1.99 & 0.53 & 0.08 \\
42--55 & 54.4 & 2.75 & 7.26 & 0.92 & 0.04 \\
55--75 & 34.7 & 1.70 & 7.12 & 1.25 & 0.07   
\label{tbl_THzabsorption}
\end{longtable}

\clearpage
\begin{longtable}{clccc}
\caption{Molecular and Atomic Abundances in the Diffuse Cloud Toward W51}\\
            & $N({\rm X})$ & $v_{\rm LSR}$  & FWHM & \\
Species & (cm$^{-2}$) & (km~s$^{-1}$)  & (km~s$^{-1}$) & References \\
\hline
\hline
\endhead
\hline
\multicolumn{5}{p{6.0in}}{{\bf Notes:} Values of $v_{\rm LSR}$ and FWHM describe the range of velocities over which the corresponding column densities were determined.  In cases where $v_{\rm LSR}$ and FWHM are each given by a single value, they represent the line center and full width at half maximum for a Gaussian fit to the absorption feature.  In cases where $v_{\rm LSR}$ is given by a range, the column density was determined by integrating between these two velocities.  The H$_2$O$^+$ column density was determined from our observations of {\it ortho}-H$_2$O$^+$ and an assumed {\it ortho}-to-{\it para} ratio of 3. \newline
{\bf References:} (1) Calculated from the H~\textsc{i} absorption line parameters for G49.5--0.4e tabulated in \citet{koo1997}, but rescaled for a spin temperature of 100~K; (2) estimated from $N({\rm CH})$ reported by \citet{gerin2010ch} and the empirical relationship between CH and H$_2$ reported by \citet{sheffer2008}; (3) this work; (4) \citet[][in press]{gerin2012}; (5) \citet{gerin2010ch}
}
\endlastfoot
H                           & $(1.39\pm0.3)\times10^{21}$   &     6.2 &   5.6 & 1 \\
H$_2$                   & $(1.06\pm0.52)\times10^{21}$ & 3--10 &     ... & 2 \\
H$_3^+$               & $(2.89\pm0.37)\times10^{14}$ &     5.0 & 10.5 & 3 \\
OH$^+$                & $(2.97\pm0.13)\times10^{13}$ & 0--11 & ... & 3 \\
H$_2$O$^+$        & $(6.09\pm0.96)\times10^{12}$ & 0--11 & ... & 3 \\
C$^+$                  & $(4.0\pm0.4)\times10^{17}$     & 0--11 & ... &  4 \\
CH                         & $(3.7\pm0.2)\times10^{13}$    & 3--10 &     ... & 5 \\
CO                        & $(2.81\pm0.21)\times10^{15}$ &      4.8 & 9.6 & 3
\label{tbl_allmol}
\end{longtable}

\end{document}